\documentclass[aps,preprint,noshowpacs,superscriptaddress,groupedaddress]{revtex4}  
\usepackage{graphicx}  
\usepackage{dcolumn}   
\usepackage{bm}        
\usepackage{amssymb}   
\usepackage{color}

\begin{document}

\title{Ultrasensitive force detection with a nanotube mechanical
resonator}

\author{J. Moser}
\affiliation{ICFO, Av. Carl Friedrich Gauss, 08860 Castelldefels,
Barcelona, Spain}\affiliation{ICN, CIN2-CSIC, Campus UAB, 08193
Barcelona, Spain}

\author{J. G\"uttinger}
\affiliation{ICFO, Av. Carl Friedrich Gauss, 08860 Castelldefels,
Barcelona, Spain}\affiliation{ICN, CIN2-CSIC, Campus UAB, 08193
Barcelona, Spain}

\author{A. Eichler}
\affiliation{ICFO, Av. Carl Friedrich Gauss, 08860 Castelldefels,
Barcelona, Spain}\affiliation{ICN, CIN2-CSIC, Campus UAB, 08193
Barcelona, Spain}

\author{M. J. Esplandiu}
\affiliation{ICN, CIN2-CSIC, Campus UAB, 08193 Barcelona, Spain}

\author{D. E. Liu}
\affiliation{Department of Physics and Astronomy, Michigan State
University, East Lansing, Michigan 48824, USA}

\author{M. I. Dykman}
\affiliation{Department of Physics and Astronomy, Michigan State
University, East Lansing, Michigan 48824, USA}

\author{A. Bachtold}
\thanks{Corresponding author: adrian.bachtold@icfo.es}
\affiliation{ICFO, Av. Carl Friedrich Gauss, 08860 Castelldefels,
Barcelona, Spain}\affiliation{ICN, CIN2-CSIC, Campus UAB, 08193
Barcelona, Spain}

\begin{abstract}
\textbf{Since the advent of atomic force microscopy
\cite{Binnig1986}, mechanical resonators have been used to study a
wide variety of phenomena, such as the dynamics of individual
electron spins \cite{Rugar2004}, persistent currents in normal
metal rings \cite{Harris2009}, and the Casimir force
\cite{Mohideen1998,Capasso2001}. Key to these experiments is the
ability to measure weak forces. Here, we report on force sensing
experiments with a sensitivity of 12~zN/$\sqrt{\textrm{Hz}}$ at a
temperature of 1.2~K using a resonator made of a carbon nanotube.
An ultra-sensitive method based on cross-correlated electrical
noise measurements, in combination with parametric
down-conversion, is used to detect the low-amplitude vibrations of
the nanotube induced by weak forces. The force sensitivity is
quantified by applying a known capacitive force. This detection
method also allows us to measure the Brownian vibrations of the
nanotube down to cryogenic temperatures. Force sensing with
nanotube resonators offers new opportunities for detecting and
manipulating individual nuclear spins as well as for magnetometry
measurements.}
\end{abstract}

\maketitle

Force sensing with a mechanical resonator consists in converting a
weak force $F$ into a displacement $z$ that is measurable by
electrical or optical means. Advances in microfabrication in the
late 1990's made it feasible to reach a force sensitivity of
820~zN/$\sqrt{\textrm{Hz}}$ with ultra-soft cantilevers
(1~zN$=10^{-21}$~N) \cite{Rugar1997,Rugar2001}. In spite of
intensive efforts over the last decade, progress in force
sensitivity has been modest. These efforts include using new
materials for the resonator, such as diamond \cite{Degendiamond};
improving the displacement detection
\cite{Teufel2009,Kippenberg2012}, which can reach an imprecision
below that at the standard quantum limit; and developing novel
resonators, such as optically levitated nanospheres
\cite{Chang2009,Raizen2011,Quidant2012}. Optimizing both the
resonator and its readout have led to a record sensitivity of
510~zN/$\sqrt{\textrm{Hz}}$ \cite{Teufel2009}.

A promising strategy for measuring lower forces is to employ
resonators made of a molecular system, such as a carbon nanotube
\cite{Reulet2000,Sazanova2004,Lassagne2009,Steele2009,Alex2011}.
Nanotube resonators are characterized by an ultra-low mass $M$,
which can be up to seven orders of magnitude lower than that of
the ultra-soft cantilevers mentioned above \cite{Rugar2001},
whereas their quality factor $Q$ can be high \cite{Huttel2009} and
their spring constant $k_{0}$ low. This has a great potential for
generating an outstanding force sensitivity, whose classical limit
is given by
\begin{equation}
S_{F}=4k_{B}T\gamma=4k_{B}T \sqrt{Mk_{0}}/Q.
\end{equation}
Here $k_{B}T$ is the thermal energy and $\gamma$ the mechanical
resistance \cite{Rugar2001}. This limit is set by the
fluctuation-dissipation theorem, which associates Langevin
fluctuating forces with the irreversible losses existing in a
resonator, quantified by $Q$. Such losses may originate, for
instance, from the phononic or the electronic thermal bath coupled
to the resonator. Measuring the thermal vibrations, i.e. the
Brownian motion, of the resonator demonstrates that its actual
force sensitivity is limited by the Langevin fluctuating forces.

A challenge with resonators based on nanotubes is to detect their
low-amplitude vibrations, since these vibrations are transduced
into electrical and optical signals that are small and have to be
extracted from an overwhelmingly large noise background. In
particular, the thermal vibrations of a nanotube have not been
detected below room temperature \cite{Favero2012}. The best force
sensitivity achieved thus far with nanotube resonators
\cite{Sazanova2004,Alex2011} has been limited by noise in the
electrical measurement setup, and has not surpassed the record
sensitivity obtained with other resonators.

To efficiently convert weak forces into sizable displacements, we
design nanotube resonators endowed with spring constants as low as
$\sim10$~$\mu$N/m. This is achieved by fabricating the longest
possible single-wall nanotube resonators. The fabrication process
starts with the growth of nanotubes by chemical vapor deposition
onto a doped silicon substrate coated with silicon oxide. Using
atomic force microscopy (AFM), we select nanotubes that are
straight over a distance of several micrometers, so that they do
not touch the underlying substrate once they are released
\cite{Chaste2011}. We use electron-beam lithography to pattern a
source and a drain electrode that electrically contact and
mechanically clamp the nanotube. We suspend the nanotube using
hydrofluoric acid and a critical point dryer. Figure~1a shows a
nanotube resonator that is 4~$\mu$m long. We characterize its
resonant frequencies by driving it electrostatically and using a
mixing detection method \cite{Gouttenoire2010,Alex2011}. The
lowest resonant frequency is 4.2~MHz (Fig.~1c). This gives a
spring constant of $7$~$\mu$N/m using an effective mass of
$10^{-20}$~kg, estimated from the size of the nanotube measured by
AFM (supplementary information). This spring constant is
comparable to that of the softest cantilevers fabricated so far
\cite{Rugar1997}. When changing the gate voltage $V_{g}^{DC}$
applied to the silicon substrate, the resonant frequency splits
into two branches (Fig.~1c). These two branches correspond to the
two fundamental modes; they vibrate in perpendicular directions
(inset to Fig.~1c).

We have developed an ultrasensitive detection method based on
parametric down-conversion, which \emph{(i)} employs a
cross-correlation measurement scheme to reduce the electrical
noise in the setup and \emph{(ii)} takes advantage of the high
tansconductance of the nanotube in the Coulomb blockade regime to
convert motion into a sizable electron current. Our detection
scheme, which is summarized in Fig.~2a, proceeds as follows. The
oscillating displacement of the nanotube, induced by the Langevin
fluctuating forces, modulates the capacitance $C_{g}$ between the
nanotube and the gate, which in turn yields a modulation $\delta
G$ of the conductance of the nanotube. We apply a weak oscillating
voltage of amplitude $V_{sd}^{AC}$ on the source electrode at a
frequency $f_{sd}$ a few tens of kHz away from the resonant
frequency $f_{0}$. (We verify that the amplitude of the thermal
vibrations does not change upon varying $V_{sd}^{AC}$; see
supplementary information.) The resulting current fluctuations at
the drain electrode at frequency $\sim|f_{sd}-f_{0}|$ are
described by
\begin{equation}
\delta I=V_{sd}^{AC}\delta G
=V_{sd}^{AC}\frac{dG}{dV_{g}}V_{g}^{DC}\frac{C_{g}^{\prime}}{C_{g}}\delta
z(t)\cos(2\pi f_{sd}t),
\end{equation}
where $dG/dV_{g}$ is the static transconductance of the nanotube,
and $\delta z$ the fluctuational displacement along the $z$ axis
(Fig.~1c). In order to enhance $\delta I$, we select a nanotube
that features sharp Coulomb blockade peaks (Fig.~1b), so that
$dG/dV_{g}$ is high for certain values of $V_{g}^{DC}$. We then
convert current fluctuations into voltage fluctuations across a
resistor $R=2$~k$\Omega$. This voltage signal is amplified by two
independent low-noise, high-impedance amplifiers. We perform the
cross-correlation of the output of the two amplifiers using a fast
Fourier transform signal analyzer
\cite{Glattli1997,Glattli1997b,Henny1999}. As a result, the
voltage noise of the amplifiers cancels out, while the weak signal
of the thermal vibrations can be extracted from the noise
background (see the supplementary information for details). This
procedure allows us to measure the power spectral density of
current fluctuations through the nanotube, which reads
$S_{I}=\left\langle\delta I^{2}\right\rangle/$rbw. Here, rbw is
the resolution bandwidth of the measurement and
$\left\langle\delta I^{2}\right\rangle$ is the mean square Fourier
component of the time-averaged current cross-correlation at
frequency $\sim|f_{sd}-f_{0}|$. Figures~2b,c show the resonance of
the thermal vibrations at 1.2~K for the two modes characterized
above, which are hereafter labeled mode~1 and mode~2. The
lineshapes are well described by a Lorentzian function.

We observe the coupling between thermal vibrations and electrons
in the Coulomb blockade regime by collecting $S_{I}$ spectra as a
function of $V_{g}^{DC}$ for these two modes (Figs.~3a,b). The
resonant frequency of mode~2 oscillates as a function of
$V_{g}^{DC}$ with the same period as the conductance oscillations
(Fig.~3c) while this dependence is monotonous for mode~1. As for
damping, the resonance lineshape of mode~2 is much wider than the
resonance lineshape of mode~1. This is readily seen in Figs.~2b,c,
where we measure $Q=13,000$ for mode~2 and $Q=48,000$ for mode~1.
To understand why mode~1 and mode~2 exhibit distinct features, we
recall that Coulomb blockade enhances the coupling between
vibrations and electrons in the nanotube
\cite{Lassagne2009,Steele2009,Wernsdorfer2012}, causing
oscillations in resonant frequency as well as additional
dissipation. The magnitude of both effects scales with the
modulation of $C_{g}$ induced by the nanotube vibrations, that is,
with the nanotube displacement projected onto the $z$ direction
perpendicular to the gate. The distinct behaviors measured for
modes~1 and~2 indicate that mode~1 essentially vibrates parallel
to the direction of the gate while mode~2 vibrates perpendicularly
to it (inset to Fig.~1c). The angle $\theta$ between the
vibrations of mode~1 and the direction parallel to the gate can be
estimated by comparing the integrated areas of the measured
spectra of modes~1 and~2, which also depend on $C_{g}$. This
results in $\theta=19.5^{\circ}\pm 2^{\circ}$ in the studied range
of $V_{g}^{DC}$ (see the supplementary information).

We then measure the force sensitivity of the resonator using a
calibrating force. For this, we apply a capacitive force on mode~1
of amplitude $F_{d}=C_{g}^{\prime}V_{g}^{DC}V_{g}^{AC}\sin\theta$,
with $V_{g}^{AC}(t)$ a small oscillating gate voltage at the
resonant frequency of mode~1. We perform the calibration with this
mode, since its high $Q$ leads to higher force sensitivity. As a
result, the driven vibrations appear as a sharp peak superimposed
on the thermal resonance in the $\left\langle\delta
I^{2}\right\rangle$ spectrum of mode~1 (Fig.~4a). The square root
of the height of this peak, $I_\mathrm{peak}$, scales linearly
with $V_{g}^{AC}$, as expected (Fig.~4b). By comparing the height
of the driven peak with that of the thermal resonance using
\begin{equation}
S_{F}=\frac{\textrm{thermal resonance height}}{\textrm{driven peak
height}}\times F_{d}^{2}/\textrm{rbw},
\end{equation}
we obtain $\sqrt{S_{F}}=12\pm8$~zN/$\sqrt{\textrm{Hz}}$ at
$T=1.2$~K. Here, we use
$C_{g}^{\prime}=1.2(\pm0.4)\times10^{-12}$~F/m, estimated from the
spacing in gate voltage between the Coulomb blockade peaks and the
effective distance between the nanotube and the gate. The
uncertainties in $\sqrt{S_{F}}$ reflect imprecisions in
$C_{g}^{\prime}$, $\theta$, and the heights of the driven peak and
the thermal resonance. See the supplementary information for
details on the measurement of the force sensitivity. Within the
experimental uncertainties, the measured force sensitivity is in
agreement with the value expected from the fluctuation-dissipation
theorem (Eq.~1), which is $23\pm5$~zN/$\sqrt{\textrm{Hz}}$. The
latter uncertainties reflect imprecisions in the effective mass of
the nanotube (supplementary information) and the temperature.

By raising the temperature to 3~K, the Langevin fluctuating forces
increase. Using the measurement method employed at 1.2~K, we
obtain a force sensitivity of 38~zN$/\sqrt{\textrm{Hz}}$
(Figs.~4c,d). This is in agreement with the value expected at 3~K
from
$S_{F}(3\mathrm{K})=S_{F}(1.2\mathrm{K})\frac{Q(1.2\mathrm{K})}{Q(3\mathrm{K})}\frac{3\mathrm{K}}{1.2\mathrm{K}}$
according to Eq.~1, where we use the force sensitivity measured at
1.2~K and the quality factors extracted from the resonances at 1.2
and 3~K.

Measuring the thermal vibrations of nanotube resonators sheds new
light on their dynamics. Different sources of noise in
nanomechanical resonators were discussed in Ref.~\cite{Cleland}.
Our finding that the resonance lineshape is well described by a
Lorentzian function at low temperature implies that nonlinear
damping is negligible \cite{Alex2011}, the Duffing nonlinearity is
weak, and the frequency noise is Gaussian and white \cite{Dykman}.

Carbon nanotube resonators enable an unprecedented force
sensitivity on the scale of 10~zN/$\sqrt{\textrm{Hz}}$ at 1.2~K.
We anticipate that the sensitivity will improve by at least a
factor of 10 by operating the resonator at milliKelvin
temperatures. Indeed, the quality factor of nanotube resonators is
enhanced at these temperatures \cite{Huttel2009}, so that both low
$T$ and high $Q$ reduce $S_{F}$. Nanotube resonators hold promise
for resonant magnetic imaging with single nuclear spin resolution
\cite{Rugar2004,Poggio2010,RugarPNAS,Nichol2012}. If our nanotube
resonator can be implemented in the experimental setups described
in Ref.~\cite{Rugar2004,RugarPNAS} without degrading the force
sensitivity achieved in the present work, it should be feasible to
detect a single nuclear spin \cite{Degendiamond}. A first step in
this direction will be to manipulate the nuclear spins of $^{13}$C
atoms naturally present in nanotubes using the protocol reported
in Ref.~\cite{Nichol2012}. These resonators may also be used for
ultra-sensitive magnetometery measurements of individual magnetic
nanoparticles and molecular magnets attached to the nanotube.

\vspace{20pt} \textbf{Acknowlegements}\\ We thank C.~Degen,
C.~Glattli, C.~Sch\"onenberger, J. Gabelli, and T.~Kontos for
discussions. We acknowledge support from the European Union
through the RODIN-FP7 project, the ERC-carbonNEMS project, and a
Marie Curie grant (271938), the Spanish state (FIS2009-11284), the
Catalan government (AGAUR, SGR), and the US Army Research
Office.\vspace{20pt}\\
\textbf{Author contributions}\\ J.M.
fabricated the device, developed the experimental setup, and
carried out the measurements. J.G. and A.E. provided support with
the experimental setup. J.G. participated in the measurements.
J.G. and J.M. analyzed the data. M.J.E. grew the nanotubes. D.E.L.
and M.I.D. provided support with the theory and wrote the
theoretical part of the supplementary information. J.M. and A.B.
wrote the manuscript with critical comments from J.G. and M.I.D.
A.B. conceived the experiment and supervised the work.

\newpage
\begin{center}
\begin{figure}[h]
\includegraphics{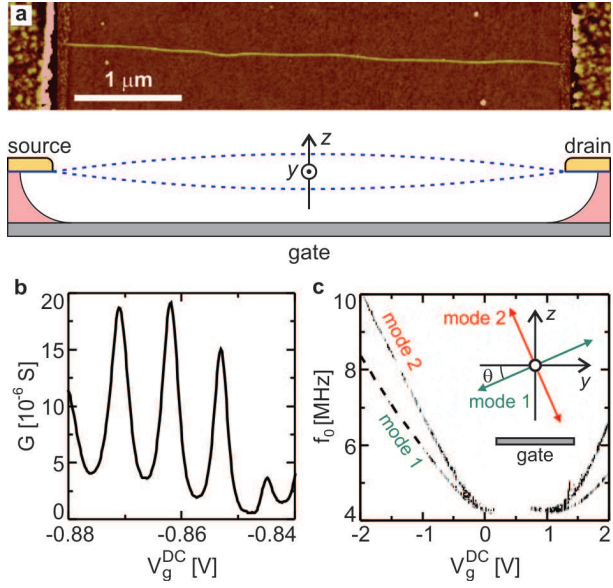}
\caption{\textbf{Nanotube resonator with low spring constant.} (a)
Atomic force microscope image of a 4~$\mu$m long nanotube prior to
removing the silicon oxide (top), and schematic of the device
(bottom). (b) Conductance $G$ of the nanotube as a function of
gate bias $V_{g}^{DC}$ at 1.2~K. (c) Resonant frequency $f_{0}$ as
a function of $V_{g}^{DC}$ in the presence of a driving force
(data obtained by measuring the mixing current with the
frequency-modulation technique \cite{Gouttenoire2010,Alex2011}).
The two lowest frequency modes are shown. We indicate the resonant
frequency of mode~1 with dashes for $V_{g}^{DC}$ ranging from -2
to -1~V, because the mixing current is weak and is difficult to
see in the figure. The resonant frequency is highly tunable, as it
can be changed by 100\% when varying $V_{g}^{DC}$ by only 1.5~V.
Inset: modes 1 and 2 vibrate along perpendicular directions;
mode~1 vibrates at an angle $\theta$ with respect to the $y$
direction, which runs parallel to the gate.}
\end{figure}
\end{center}

\newpage
\begin{center}
\begin{figure}[h]
\includegraphics{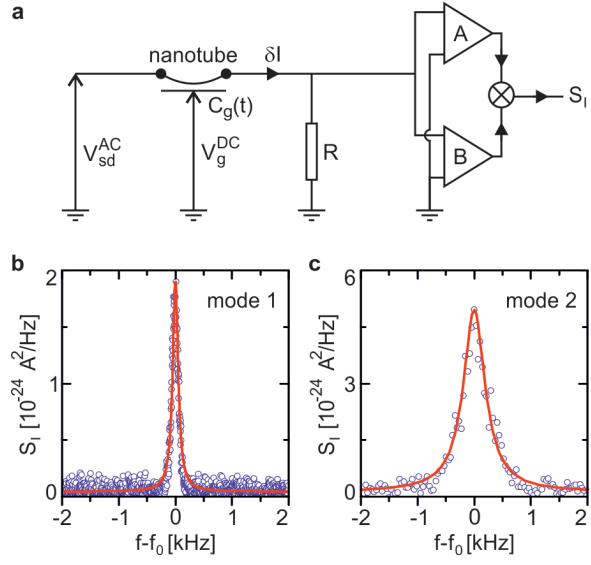}
\caption{\textbf{Measuring thermal vibrations.} (a) Schematic of
the cross-correlation measurement setup. (b,c) Power spectral
density $S_{I}$ of current fluctuations for modes 1 and 2 at
1.2~K, centered at the mode's resonant frequency. We apply
$V_{g}^{DC}=-0.854$~V and $V_{sd}^{AC}=89$~$\mu$V. Mechanical
quality factors are $Q=48,000$ for mode~1 and $Q=13,000$ for
mode~2. We find that the quality factor of mode~2 oscillates as a
function of $V_{g}^{DC}$ between 8,000 and 20,000 with the same
period as the conductance oscillations. Since the signal is weaker
for mode~1, the resonance can be clearly resolved only over a
limited range of $V_{g}^{DC}$.}
\end{figure}
\end{center}

\newpage
\begin{center}
\begin{figure}[h]
\includegraphics{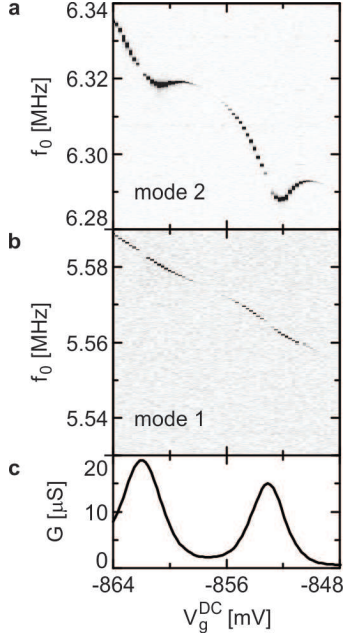}
\caption{\textbf{Electron-vibration coupling in the Coulomb
blockade regime.} (a,b) $S_{I}$ spectra showing the resonant
frequency $f_{0}$ as a function of $V_{g}^{DC}$ at 1.2~K for
modes~1 and 2. (c) Conductance $G$ as a function of $V_{g}^{DC}$
at 1.2~K. $S_{I}$ in (a,b) strongly depends on $dG/dV_{g}$, as
expected from Eq.~2. We estimate the variance of the displacement
of the thermal vibrations to be $\simeq(1.1~\textrm{nm})^2$ from
the equipartition theorem. We obtain a similar variance by
converting the $S_{I}$ spectra into displacement fluctuations.
This conversion, which depends on various parameters obtained
separately, is discussed in the supplementary information.}
\end{figure}
\end{center}

\newpage
\begin{center}
\begin{figure}[h]
\includegraphics{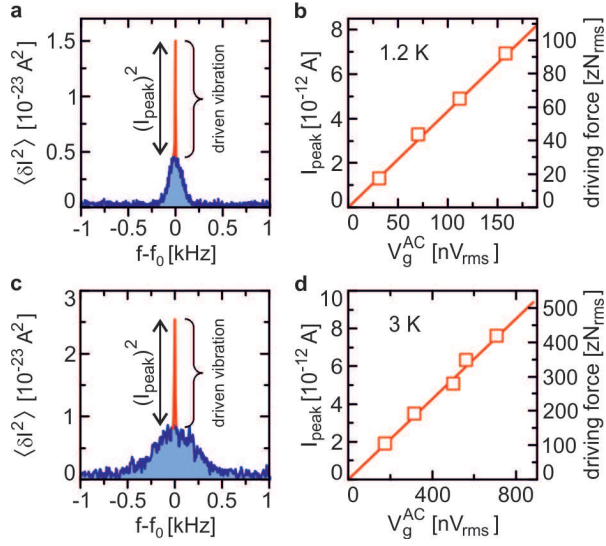}
\caption{\textbf{Force sensing experiment.} (a) The square
amplitude $\left\langle \delta I^{2}\right\rangle$ of the Fourier
transform of the current cross-correlation at 1.2~K in the
presence of a driving force at the resonant frequency of mode~1
($\left\langle \delta
I^{2}\right\rangle=S_{I}\times\mathrm{rbw}$). The driven vibration
signal is indicated in red while the thermal vibration signal is
indicated in blue. We apply $V_{sd}^{AC}=89$~$\mu$V,
$V_{g}^{DC}=-0.854$~V, and
$V_{g}^{AC}=70$~$\mathrm{nV}_{\mathrm{rms}}$, and set
$\textrm{rbw}=4.69$~Hz. (b) The square root of the driven
resonance height in (a), measured as a function of oscillating
voltage $V_{g}^{AC}$ applied to the gate. Also shown is the
driving force estimated from $V_{g}^{AC}$. (c) and (d) are
analogous to (a) and (b) at 3~K. The voltage $V_{g}^{AC}$ induces
a current of purely electrical origin (supplementary information),
whose contribution is negligible. This electrical contribution,
which can be measured with a drive off resonance, can be detected
only for exceedingly large $V_{g}^{AC}$.}
\end{figure}
\end{center}

\end{document}